\documentclass{JHEP3}
\usepackage{epsfig}

\title{Cold dark matter in brane cosmology scenario}

\author{E. Abou El Dahab\\
Basic Science Division, Modern University For Technology and
Information, Cairo,Egypt.}

\author{S. Khalil\\
Center for Theoretical Physics at the
British University in Egypt, Sherouk City, Cairo 11837, Egypt.\\
Department of Mathematics, Faculty of Science, Ain Shams
University, Cairo 11566, Egypt.}

\abstract{ We analyze the dark matter problem in the context of
brane cosmology. We investigate the impact of the non-conventional
brane cosmology on the relic abundance of non-relativistic stable
particles in high and low reheating temperature scenarios. We show
that in case of high reheating temperature, the brane cosmology
may enhance the dark matter relic density by many order of
magnitudes and a stringent lower bound on the five dimensional
scale is obtained. We also consider low reheating temperature
scenarios with chemical equilibrium and non-equilibrium. We
emphasize that in non-equilibrium case, the resulting relic
density is very small. While with equilibrium, it is increased by
a factor of ${\cal O}(10^2)$ with respect to the standard thermal
production. Therefore, dark matter particles with large cross
section, which is favored by detection expirements, can be
consistent with the recent relic density observational limits.}

\keywords{Dark Matter, Barne Cosmology}

\preprint{}

\begin{document}

\newcommand{\bmat}{\left(\begin{array}}
\newcommand{\emat}{\end{array}\right)}
\newcommand{\be}{\begin{equation}}
\newcommand{\ee}{\end{equation}}
\newcommand{\bea}{\begin{eqnarray}}
\newcommand{\eea}{\end{eqnarray}}
\def\ie{{\it i.e.}}
\def\lsim{\raise0.3ex\hbox{$\;<$\kern-0.75em\raise-1.1ex\hbox{$\sim\;$}}}
\def\gsim{\raise0.3ex\hbox{$\;>$\kern-0.75em\raise-1.1ex\hbox{$\sim\;$}}}
\def\Frac#1#2{\frac{\displaystyle{#1}}{\displaystyle{#2}}}

\section{{\large \bf Introduction}}

The history of the observed Universe is not known before the epoch
of nucleosynthesis (\ie, temperature above $1$ MeV). The
assumption that our universe has to be homogeneous and isotropic
is only true at late universe. Therefore, one can consider the
standard cosmology based on Robertson-Walker (RW) metric as an
effective low energy theory for a more fundamental one at higher
energy scale.

Warped extra dimensions have been proposed~\cite{Randall:1999ee}
to explain the large hierarchy between the electroweak scale ($M_W
\sim 10^2$ GeV) and the Planck scale ($M_{pl} \sim 10^{19}$ GeV).
In this class of models, the ordinary matter is assumed to be
localized on a three-dimensional subspace, called brane which is
embedded in a larger space, called bulk. It has been emphasized
that the brane cosmology in these models can be quite different
from the standard cosmology of four dimensional universe. In
particular, the derived Friedman equation of a brane embedded in
five dimensional ($5D$) warped geometry is given by
\cite{Binetruy:1999ut} \be H^2 = \frac{8 \pi G_{(4)}}{3} \rho
\left( 1 + \frac{\rho}{2 \sigma}\right) - \frac{k}{a^2}+
\frac{{\cal C}}{a^4}, \ee where $H=\dot{a}/a$ is the Hubble
parameter and $a(t)$ is the scale factor. $\rho$ is the energy
density of ordinary matter on the brane while $\sigma$ is the
brane tension. $G_{(4)}$ refers to the $4D$ Newton coupling
constant. Finally $k$ stands for the curvature of our three
spatial dimensional and ${\cal C}$ is a constant of integration
known as dark-radiation. As can be seen from the above equation,
$H \propto \rho$ rather than $\sqrt{\rho}$ as in the conventional
cosmology. Thus, the evolution of the scale factor will be
different from the standard one. This modification would be very
relevant for the cosmological events, as dark matter relic
abundance, that may occur during the radiation dominated phase of
the early universe.

One of the theoretical problem in modern cosmology is the
existence of dark matter (DM). The most interesting candidate for
this dark matter is a long lived or stable weakly interacting
massive particle (WIMP) which can remain from the earliest moments
of the universe in sufficient number to account for the dark
matter relic density. The standard computation of the WIMP relic
density, based on the usual early universe assumptions, leads to
the following fact: The WIMP relic density is inversely
proportional with its annihilation cross section. However, the
detection rate of this particle, which is given in terms of its
elastic cross section with the nuclei in the detector, is
proportional to its annihilation cross section. Therefore, in
order to detect the WIMP experimenally, its cross section should
large, of ${\cal O}(10^{-6}- 10^{-8})\rm{GeV}^{-2}$. Neverthless,
in this case the WIMP relic density is quite small: $\Omega_{\chi}
h^2 \lsim 0.01$ which contradicts the recent observational bounds:
$0.094 \lsim \Omega_{\chi} h^2 \lsim 0.128$ \cite{Bennett:2003bz}.

The aim of this paper is to study the relic density in the context
of non-conventional brane cosmology. It is important to mention
that there are some attempts for similar analysis in the
literature \cite{Okada:2004nc,Nihei:2005qx}, however, that work
was focused on the standard thermal production scenario for
computing the relic density. In this case, the reheating
temperature is assumed to be much larger than the freeze-out
temperature and the dark matter particles are in thermal
equilibrium. Here, we analyze the WIMP relic density in
non-conventional brane cosmology in case of high and low reheating
temperature. We show that in the first scenario where the
reheating temperature is high, a strong lower bound is imposed on
the $5D$ fundamental scale. While in case of low reheating
temperature and the WIMPs reach a chemical equilibrium before
reheating, the relic density is enhanced by one-two order of
magnitudes. This enhancement could help to accommodate large
annihilation cross section DM which, as mentioned, has a very
small relic density in the standard scenario. As an example, we
apply our result for the lightest supersymmetric particle (LSP)
which is one of the best candidate for cold dark matter.

The paper is organized as follows. In section 2 we briefly review
the non-conventional brane cosmology and point out that the
consistency of the boundary conditions on the brane implies that
the $5D$ metric should contain two warp factors at least,
otherwise the relation $\rho_b =- p_b$ is identically satisfied.
Section 3 is devoted for analyzing the WIMP relic density in brane
cosmology when the reheating temperature is much higher than the
decoupling temperature. We show that in this case the relic
density is enhanced by many order of magnitudes which set a strong
constraint on $M_5$. In section 4 we study the DM relic density in
brane cosmology with low reheating temperature where DM particles
could be in chemical equilibrium or non-equilibrium. We show that
the former case is the most interesting scenario, since the relic
abundance is increased by a factor of ${\cal O}(10^2)$. Therefore
it can be an interesting possiblity for accommodating the recent
relic density observational limits with large cross sections .
Finally we give our conclusions in section 5.

\section{\large{\bf Brane world cosmology}}

In this section we discuss the effective brane cosmology from
Einstein equations in five dimensional space-time. We consider the
most general metric that preserves three dimensional rotational
and translation invariance: \be ds^2 = - n^2 ( t, y ) dt^2 + a^2 (
t, y ) \gamma_{ij} dx^i dx^j + b^2(t,y) dy^2, \label{metric}\ee
where $\gamma_{ij}$ is a maximally symmetric 3-dimensional metric
with spatial curvature $k=\pm 1,0$. In our analysis, we identify
the hypersurface defined by $y=0$ with the brane that forms our
universe. The induced metric in this brane is the usual $4D$ RW
metric.

The five-dimensional Einstein equations (with bulk cosmological
constant $\Lambda$) are given by \be \tilde{R}_{AB}
-\frac{1}{2}\left(\tilde{R} - 2 \Lambda\right) \tilde{g}_{AB} =
\kappa_{( 5 )}^2 \tilde{T}_{AB}, \label{GR}\ee where
$\tilde{R}_{AB}$ is the $5D$ Ricci tensor, $\tilde{R}$ is the $5D$
scalar curvature and the constant $\kappa_{(5)}$ is related to the
$5D$ Newton's constant $G_{(5)}$ and the $5D$ Plank mass
$M_{(5)}$, by the relations \be \kappa_{(5)}^2 = 8 \pi G_{(5)} =
M^{-3}_{(5)} . \ee

Assuming an empty bulk, the energy momentum tensor is due to the
matter on the brane, which is considered to be an infinitely thin.
Thus $T_A^{~B}$ is given by \be T_A^{~B} = \frac{\delta ( y )}{b}
\rm{diag} \left(-\rho_b, p_b, p_b, p_b,0\right), \ee where
$\rho_b$ and $p_b$ are the total energy density and pressure on
the brane, respectively. In order to have a well defined geometry,
the metric (\ie, the function $f \equiv n,a,b$ in
Eq.(\ref{metric})) must be continuous across the brane localized
at $y=0$ but its derivative with respect to $y$ can be
discontinuous. Therefore, its second derivative with respect to
$y$ will contain Dirac delta function, \ie, \be f^{''} =
\widehat{f^{''}} +[f^{'}] \delta(y),\ee where $\widehat{f^{''}}$
is the non-distributional part of the second derivative of $f$
respect to $y$ and $[f^{'}]$ is the jump in the first derivative
of $f$ across $y=0$ which is defined as \be [f^{'}] = f^{'}(0^+)
-f^{'}(0^-).\ee By matching the Dirac $\delta(y)$ function in
Eq.(\ref{GR}), one obtains the following relations, which are
known as junction conditions \cite{Binetruy:1999ut}: \bea
\frac{[a^{'}]}{a_0 b_0} &=& - \frac{\kappa^2_{(5)}
\rho_b}{3},\label{jun1}\\
\frac{[n^{'}]}{n_0 b_0} &=& \frac{\kappa^2_{(5)} (2 \rho_b+ 3
p_b)}{3}, \label{jun2}\eea  where the subscript $0$ stands for the
evaluation at $y = 0$. These conditions imply that  the $5D$
metric should contain at least two different warp factors,
otherwise we get $p_b= - \rho_b$ as an identical relation which
means that our universe is always vacuum dominated. This is not
consistent and contradicts with the assumption of having matter on
the brane. For instance, if we consider the case of embedding the
$4D$ RW metric in the following $5D$ metric: \be ds^2 =
h^2(t,y)\left[-dt^2 + R(t)^2 \gamma_{ij} dx^i dx^j \right] +
b^2(t,y) dy^2. \ee In this case, one finds that $a^2(t,y)\equiv
R(t)^2 h^2(t,y)$, $n^2(t,y)=h^2(t,y)$, and the junction conditions
(\ref{jun1}),(\ref{jun2}) lead to the relation \be p_b =- \rho_b
,\ee \ie, the vacuum energy is always dominated on the brane,
which is not necessary true. It is worth mentioning that this
result does not depend on the choice of the scale factor $b(t,y)$
of the fifth dimension or on our cosmological scale factor
$R(t)$\footnote{For a similar observation in brane cosmology of
Jordon-Brans-Dick theory, see Ref. \cite{Arik:2005gf}.}.

Using the condition (\ref{jun1}) with the components $(0,0)$ and
$(4,4)$ of Einstein's equations in the bulk, one finds the
following equation \be H^2 \equiv \frac{\dot{a}_0^2}{a_0^2} =
\frac{\Lambda}{6} +
   \frac{\kappa_{(5)}^4}{36} \rho_b ^2 - \frac{k}{a^2}+ \frac{{\cal C}}{a^4}, \ee
where ${\cal C}$ is a constant of integration. As can be seen from
this equation, the Hubble parameter is proportional to the energy
density of the brane, in contrast with the standard
four-dimensional Friedmann equation where it depends on the square
root of the energy density. Let us consider a brane with total
energy density \be \rho_b = \sigma + \rho, \ee where $\sigma$ is a
brane tension, constant in time, and $\rho$ is the energy density
of ordinary cosmological matter. This implies \be H^2 = \left(
\frac{\kappa_{(5)}^4 \sigma^2}{36} + \frac{\Lambda}{6}\right) +
\frac{\kappa_{(5)}^4}{18} \sigma \rho +
\frac{\kappa_{(5)}^4}{36}\rho^2 - \frac{k}{a^2}+ \frac{{\cal
C}}{a^4}. \label{EqH}\ee Then with fine tuning of brane tension,
the first term in Eq.(\ref{EqH}) vanishes if we have \be
\frac{\kappa_{( 5 )}^4}{36} \sigma^2 = -
   \frac{\Lambda}{6}. \ee
Furthermore, fixing the value of $\Lambda$ in terms of $M_{(5)}$
and $M_{pl}$ as $\Lambda = - 6 M_{(5)}^6/M_{pl}^4$, one finds \be
8 \pi G_{( 4 )} = \frac{\kappa_{(5)}^4}{6}\sigma, \ee which leads
to the new Friedmann equation: \be H^2 = \frac{8 \pi G_{(4)}}{3}
\rho \left( 1 + \frac{\rho}{2 \sigma}\right) - \frac{k}{a^2}+
\frac{{\cal C}}{a^4}.\label{Friedman} \ee As can be seen from
Eq.(\ref{Friedman}), at low energies \ie, at late time, the
cosmology can be reduced to the standard one, but in the early
time the $\rho^2$ term becomes dominant, so the universe undergoes
nonconventional cosmology.

Before concluding this section, it is worth mentioning that the
normal energy conservation equation is still valid, \be
  \dot{\rho} + 3 H ( \rho + p ) =0,
\ee which implies that $\rho \propto a_0^{-3(1+w)}$. Using this
relation in Friedmann equation (\ref{Friedman}) one finds
\cite{Binetruy:1999ut} \be a_0(t) \propto t^q, \ee where \be q
=\Big\{^{\frac{1}{3(1+w)},~~~~~~~ \rho \gg
\sigma}_{\frac{2}{3(1+w)}~~~~~~~~ \rho \ll \sigma}.\ee  Therefore,
within the radiation domination epoch where $w= 1/3$ one gets
$a_0(t) \propto t^{\frac{1}{4}}$ in the non-conventional cosmology
limit and $a_0 ( t ) \propto t^{\frac{1}{2}}$ in the standard case
limit. Thus the time dependence of Hubble parameter in raditation
domination era is given be \be H = \frac{q}{t}, ~~~~~~~~~~~~~~~~~
{\rm with}~ q =\Big\{^{\frac{1}{4},~~~~~~~~~~~~~~ \rho \gg
\sigma}_{\frac{1}{2}~~~~~~~~~~~~~~~\rho \ll
\sigma}.\label{H(t)}\ee As mentioned in the introduction, one
would expect important impact for these modifications on the
cosmological events that occurred in the early universe during the
radiation domination era. In the next section we will analyze the
effect of the non-conventional cosmology on the dark matter relic
abundance.

%
\section{\large{\bf DM relic density in non-conventional brane cosmology}}
In this section we compute the relic density of the WIMP $(\chi)$
within the non-conventional brane cosmology.  In the standard
computation for the WIMP relic density, one assumes that $\chi$
was in thermal equilibrium with the standard model particles in
the early universe and decoupled when it was non-relativistic.
Once the $\chi$ annihilation rate $\Gamma_{\chi} =\langle
\sigma^{ann}_{\chi}\ v \rangle n_{\chi}$ dropped below the
expansion rate of the universe, $\Gamma_{\chi} \leq H$, the WIMPs
stop to annihilate, fall out of equilibrium and their relic
density remains intact till now . The above $\langle
\sigma^{ann}_{\chi}\ v \rangle$ refers to thermally averaged total
cross section for annihilation of $\chi \chi$ into lighter
particles times the relative velocity, $v$.

The relic density is then determined by the Boltzmann equation for
the WIMP number density $(n_{\chi})$ and the law of entropy
conservation: \bea \frac{d n_{\chi}}{dt} &=& -3 H n_{\chi}  -
\langle \sigma^{ann}_{\chi}\ v \rangle \left[(n_{\chi})^2 -
(n^{eq}_{\chi})^2\right],\\
\frac{d s}{dt} &=& - 3 H s, \label{S(T)}\eea where $n_{\chi}^{eq}$
is the WIMP equilibrium number density which, as function of
temperature $T$, is given by $n^{eq}_{\chi} = g_{\chi} (m_{\chi}
T/2\pi)^{3/2} e^{-m_{\chi}/T}$. Here $m_{\chi}$ and $g_{\chi}$ are
the mass and the number of degrees of freedom of the WIMP
respectively. Finally, $s$ is the entropy density. In the standard
cosmology, the Hubble parameter $H$ is given by $H^2 = \left(8
\pi/3 M^2_{pl}\right) \rho$ to be compared with the expression in
Eq.(\ref{Friedman}) for brane cosmology. In our analysis we will
set $k={\cal C}=0$ in order to focus on the impact of the
modification of the $\rho$ dependence in $H$.

Let us introduce the variable $x =m_{\chi}/T$ and define $Y
=n_{\chi}/s$ with $Y_{eq} =n^{eq}_{\chi}/s$. In this case, the
Boltzmann equation is given by  \be \frac{dY}{dx} = \frac{1}{3H}
\frac{d s}{dx} \langle \sigma^{ann}_{\chi}\ v \rangle \left(Y^2 -
Y^2_{eq}\right).\ee In radiation domination era, the entropy, as
function of the temperature, is given by $ s= \frac{2\pi^2}{45}
g_{\ast_s}(x)~ m_{\chi}^3~ x^{-3}\equiv k_1 x^{-3}$, which is
deduced from the fact that $s=(\rho+ p)/T$ and $g_{\ast_s}$ is the
effective degrees of freedom for the entropy density. Therefore
one finds \be \frac{d s}{d x} = - \frac{3 s}{x},\ee which is the
same in both cases of standard and brane cosmology. It is worth
mentioning that since the time variation of $s$ is given in terms
of the Hubble parameter $H$, $s(t)$, unlike $s(T)$, is expected to
be different in brane cosmology from the usual expression in the
standard case. This indicates that the time-temperature relation
is modified in brane cosmology. As known in the standard cosmology
this relation is given by \be t_{s} = \frac{1}{2}~\sqrt{\frac{45
M_{pl}^2}{4 \pi^3 g_*}}~ T^{-2}. \ee In brane cosmology with
dominant $\rho^2$ term in the Hubble parameter $H$, the
time-temperature relation takes the following form \be t_{b} =
\frac{1}{4}~\sqrt{\frac{32400 M_5^6}{\pi^4 g_*^2}}~ T^{-4}. \ee
This means that within the brane cosmology, the cooling of the
universe becomes much slower than it in standard cosmology. In the
standard case, the following expression for the Boltzmann equation
for the WIMP number density is obtained \be \frac{dY}{dx} = -
\frac{s}{H x} \langle \sigma^{ann}_{\chi}\ v \rangle \left(Y^2 -
Y^2_{eq}\right),\label{Ygeneral}\ee where $H(x)$ is given by $ H =
\sqrt{{\frac{4\pi^3 g_\ast m_{\chi}^4}{45M^2_{pl}}}} x^{-2}=
\sqrt{k_2} x^{-2}$ and $g_*$ is the effective degrees of freedom
for the energy density. Therefore, one obtains the following
expression for the Boltzmann equation for the WIMP number density
($g_{*s} \simeq g_*$ is assumed): \be
\left(\frac{dY}{dx}\right)_{s}= -\sqrt{\frac{\pi g_*}{45}} M_{pl}~
m_{\chi} \frac {\langle \sigma^{ann}_{\chi}\ v \rangle }
{x^2}\left(Y^2 - Y^2_{eq}\right).\label{Boltzmann1}\ee

In brane cosmology the Hubble parameter is given by $H= (k_2
x^{-4} + k_3 x^{-8})^{1/2}$ where $k_{3} = \pi^4 g_\ast ^2
m_{\chi}^8/(32400 M_5^6)$. Thus, the Boltzmann equation in brane
cosmology takes the form \be \left(\frac{dY} {dx}\right)_b =
-\sqrt{\frac{\pi g_*}{45}} M_{pl}~ m_{\chi}~ {\left(x^{4} +\frac{
k_{3}}{k_2}\right)^{-1/2}}~ \langle \sigma^{ann}_{\chi}\ v \rangle
\left(Y^2-Y^2_{eq}\right). \label{Boltzmann2}\ee

It is worth noticing that in the limit of $k_3\to 0$ (\ie, $\sigma
\to \infty$), the above equation tends to the standard Boltzmann
equation in Eq.(\ref{Boltzmann1}). Therefore, at early times, the
universe undergoes a nonstandard brane cosmology till it reaches a
temperature, known as transition temperature $T_t$ where the
universe sustains the standard cosmology. This transition
temperature is defined as \cite{Okada:2004nc} \be \rho ({T_t})=
2\sigma ~~ \Rightarrow ~~ T_t = 0.51 \times 10^{-9} M_5^\frac{3}
{2}~\rm{GeV}\ee

In order to analyze the brane cosmology effect on the WIMP relic
density, one should assume that the freeze out temperature of the
WIMP $(T_F)$ is higher than the transition temperature, \ie, $T_F
\geq T_t$. Therefore, one finds \be M_5 \leq 1.57 \times 10^6
\left(\frac{m_{\chi}}{x_F}\right)^{2/3} \label{upperbound}.\ee
Since the WIMPs freeze out at temperature $T_F \ll m_{\chi}$, they
are non-relativistic and therefore the averaged annihilation cross
section can be expanded as follows: \be \langle
\sigma^{ann}_{\chi} v \rangle = a + \frac{6b} {x},\ee where $a$
describes the $s$-wave annihilation and $b$ comes from both $s$-
and $p$- wave annihilation. To obtain the present WIMP abundance
$Y_{\infty}$, we should integrate the Boltzmann equation for the
WIMP number density from $x_F$ (the decoupling temperature) to
$x_{\infty}\simeq \infty$ (present temperature). It is important
to notice that this integral must be divided to two parts from
$x_F$ to $x_t$ where the non-conventional brane cosmology is
applied and from $x_t$ to $\infty$ where the universe undergoes
the standard cosmology. Thus, one obtains \be Y^{-1}_{\infty b}=
\sqrt{\frac{\pi g_\ast}{45}}M_{pl} m_{\chi} \left[
\int^{x_t}_{x_F} (a + \frac{6b}{x}) \left(\frac{k_3}{k_2} +
x^4\right)^{-1/2} dx + \int^{\infty}_{x_t} \left(\frac{a}{x^2} +
\frac{6b}{x^3}\right)dx \right].\ee Here we have used the usual
assumption that $Y_{eq}\ll Y$ and $Y_{x_F}\gg Y_\infty$.
Evaluating the above integrals, one finds \bea Y^{-1}_{\infty b}
&=& \sqrt{\frac{\pi g_\ast}{45}}M_{pl}~ m_{\chi}
\left[3\sqrt{\frac{k_2}{k_3}} b
\left(\sinh^{-1}\left(\sqrt{\frac{k_3}{k_2}} x_F^{-2}\right) -
\sinh^{-1}\left(\sqrt{\frac{k_3}{k_2}}
x_t^{-2}\right)\right)\right. \nonumber\\
&+&\left. a \left(\frac{1}{x}~ _2F_1\left[\frac{1}{4},
\frac{1}{2}, \frac{5}{4}, \frac{-k_3}{k_2 x^4}
\right]\right)^{x_t} _{x_F} + \left(\frac{a}{x_t}+
\frac{3b}{x^2_t}\right)\right], \label{Yb-1}\eea where
$_{2}F_{1}[a,b,c,z]$ is the Hypergeometric function, which is a
solution of the hypergeometric differential equation: $z(1-z)
y^{''} +[c-(a+b+1)z] y^{'} - a b y=0$. As can be seen from
Eq.(\ref{Yb-1}) that for $x_t=x_F$ the expression of $Y_{\infty
b}^{-1}$ coincides with the standard known result for $Y_{\infty~
s}^{-1}$, namely $Y^{-1}_{\infty s}= \sqrt{\frac{\pi g_\ast}{45}}~
M_{pl}~ m_{\chi} \left(\frac{a}{x_F}+ \frac{3 b}{x^2_F}\right)$.
The relic abundance of the WIMP is given by \be \Omega_{\chi} h^2
= \frac{\rho_{\chi}}{\rho_c/h^2} = 2.9 \times 10^{8}~ Y_{\infty}
\left(\frac{m_{\chi}}{\rm{GeV}}\right),\label{omega}\ee where the
critical density $\rho_c$ is given by $\rho_c \simeq 10^{-5} h^2
~\rm{GeV} \rm{cm}^{-3}$ and $h$ is the Hubble constant, $h\simeq
0.7$. Furthermore, the $\rho_{\chi}$ is defined as $\rho_{\chi}=
m_{\chi} s_0 Y_{\infty}$ where $s_0 ~\simeq 2900~ \rm{cm}^{-3}$ is
the present entropy density. As in the standard scenario, the
relic density of the WIMP is inversely proportional to its
annihilation cross section. However, unlike the standard case, it
depends explicitly on WIMP mass since $k_3/k_2 \propto
m_{\chi}^4$.

The freeze-out temperature can be determined from the freeze-out
condition: \be \Delta (x_F) = c Y^{eq}_{\chi}(x_F), \ee where
$\Delta$ is given by $\Delta = Y_{\chi} - Y_{\chi}^{eq}$ and $c$
is a constant of order unity. In standard cosmology, this
condition implies that \be x_F = \ln \frac{0.0765~ c~ m_{\chi}
M_{pl} g_{\chi} (a+ 6b/x_F)}{\sqrt{x_F g_*(x_F)}},\ee which can be
solved iteratively to determine the value of $x_F$. It turns out
that for $m_{\chi} \sim {\cal O}(100)$ GeV, $x_F \sim {\cal
O}(25)$. In brane cosmology, one can easily show that $x_F$ is
obtained by iterative solution of \be x_F = \ln \frac{0.0765~ c~
m_{\chi} M_{pl} g_{\chi} x_F^{3/2} (a+ 6b/x_F)}{\sqrt{g_*(x_F)
(\frac{k_3}{k_2} +x_F^4)}}.\ee In this case, one finds that $x_F$
is smaller than the above value obtained within the standard
cosmology. Also, it turns out that $x_F$ is sensitive to the scale
$M_5$. For example with $M_5 \sim 10^6$ one gets $x_F \sim {\cal
O}(7)$.

Let us introduce the factor $R = (\Omega_{\chi}h ^2)_b/(\Omega
_\chi h^2)_s$ that measures the enhancement/suppression in the
relic abundance due to the brane cosmology. From Eq.(\ref{omega}),
one finds \be R = \frac{\frac{a}{x_F}+
\frac{3b}{x_F^2}}{3\sqrt{\frac{k_2}{k_3}}b\left(\sinh^{-1}\left(\sqrt{\frac{k_3}{k_2}}
x_F^{-2}\right) -
\sinh^{-1}\left(\sqrt{\frac{k_3}{k_2}}x_t^{-2}\right)\right)+
a\left(\frac{1}{x}~ _2F_1\left[\frac{1}{4},
\frac{1}{2},\frac{5}{4}, \frac{-k_3}{k_2 x^4}\right]\right)^{x_t}
_{x_F} + \frac{a}{x_t}+ \frac{3b}{x^2_t}}.\ee
This ratio could be larger or smaller than one depending on the
values of the annihilation cross section parameters $a$ and $b$
and also on the masses $m_{\chi}$ and $M_5$. In order to analyze
the impact of the non-conventional brane cosmology on the relic
density result, we consider, as an example, the lightest
supersymmetric particle (LSP), which is one of the most
interesting candidates for the WIMP. As is well known, in most of
the parameter space of the supersymmetric models the LSP is mainly
pure Bino. Therefore, it is mainly annihilated into lepton pairs
through $t-$channel exchange of right-handed sleptons. The p-wave
dominant cross section is given by \cite{Giudice:2000ex} \be b
\simeq 8 \pi \alpha_1^2 \frac{1}{m_{\chi}^2}
\frac{1}{(1+x_{\tilde{l}_R})^2}, \ee where
$x_{\tilde{l}_R}=m^2_{\tilde{l}_R}/m^2_{\chi}$ and $\alpha_1$ is
the coupling constant for the $U(1)_Y$ interaction. Thus, for
$m_{\chi} \sim m_{\tilde{l}_R}\sim 100$ GeV, one finds $b \simeq
{\mathcal O}(10^{-8})~ \rm{GeV}^{-2}$, which in the standard
cosmology scenario leads to $\Omega_{\chi} h^2 \geq 0.1$.
%
\begin{figure}[t]
\begin{center}
\epsfig{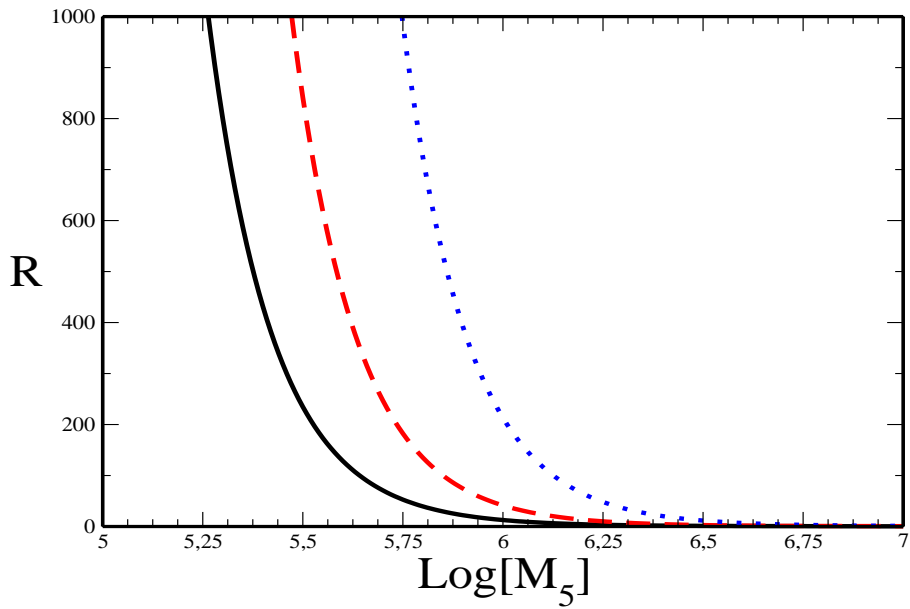}
\end{center}\caption{The enhancement/suppression factor $R=(\Omega_{\chi}h^2)_b/(\Omega_{\chi} h^2)_s$ as a function
of the five dimensional scale $M_5$ (GeV) for $m_{\chi} =100$
(solid curve), $200$ (dashed curve) and $500$ GeV (dotted curve).}
\label{RM5}
\end{figure}

As advocated above, in brane cosmology the relic density
$(\Omega_{\chi} h^2)_b$ is quite sensitive to the value of the
fundamental scale $M_5$ which should satisfy the upper bound given
in Eq.(\ref{upperbound}). Therefore, with $m_{\chi} \simeq {\cal
O}(100)$ GeV and $x_F \simeq {\cal O}(10)$, one finds \be M_5 <
10^{7}.\ee Furthermore, the fact that the transition process from
non-conventional cosmology to convention cosmology should take
place above the nucleosynthesis era (\ie, $T_t > 1$ MeV) impose
the following lower bound on $M_5$: \be M_5 \gsim 1.2 \times 10^4.
\ee

In Fig. \ref{RM5}, we present the prediction for the factor $R$ as
a function of the scale of the five dimensions, $M_5$, for
different values of $m_{\chi}$, namely we consider $m_{\chi} =
100, 200$ and $500$ GeV. As can be seen from this figure, for $M_5
< 10^6$ the brane cosmology effect is quite large and the factor
$R$ becomes much larger than one. In this case the resulting relic
density $(\Omega_{\chi} h^2)_b$ may exceed the WMAP results (at
$95\%$ confidence level) \cite{Bennett:2003bz} \be \Omega_{\chi}
h^2 = 0.1126^{+0.0161}_{-0.0181}.\ee Moreover for $M_5 \gsim 5
\times 10^{6}$, the ratio $R$ becomes less than one and a small
suppression for $(\Omega_{\chi} h^2)_s$ can be obtained. This
brane enhancement or suppression for the dark matter relic density
could be favored or disfavored based on the value of the relic
abundance in the standard scenario. If $(\Omega_{\chi}h^2)_s$ is
already larger than the observational limit, as in the case of
bino-like particle, then a suppression effect would be favored and
hence $M_5$ is constrained to be larger than $5 \times 10^{6}$
GeV. However, for wino- or Higgsino-like particle where the
standard computation usually leads to very small relic density,
the enhancement effect will be favored and the constraint on $M_5$
can be relaxed a bit \cite{Nihei:2005qx}. In general, it is
remarkable that in this scenario the dark matter relic density
imposes a stringent constraint on the fundamental scale $M_5$.

%
\section{\large{\bf DM relic density in brane cosmology with low reheating temperature}}
In the standard computation for the DM relic density that we have
adopted in the previous section, it was assumed that the reheating
temperature $T_{RH}$ is much higher than the WIMP freeze-out
temperature \ie, $T_{RH} \gg T_F$. In this case, the reheating
epoch has no impact on the final result of the relic density.
However, it is well known that the only constraint on $T_{RH}$ is
$T_{RH} \gsim 1$ MeV in order not to spoil the successful
predictions of the big bang neucleosynthesis. Therefore, in
principle it is possible to have a cosmological scenario with low
reheating such that $T_{RH} < T_F$. In this case the predictions
of the relic abundance of the WIMP are modified as emphasized in
Ref.\cite{Giudice:2000ex} and recently in
Ref.\cite{Khalil:2002mu}.

As in the previous section and to emphasize the effect of the
brane cosmology, we assume that the WIMP freeze-out temperature is
larger than the transition temperature which is also larger than
the reheating temperature, \ie, $T_F > T_t > T_{RH}$. Within a low
reheating temperature scenario, the relic density depends on
whether WIMPs are never in chemical equilibrium either before or
after reheating or they reach chemical equilibrium but they
freeze-out before the completion of the reheat process. The
resultant relic density in these two scenarios are quite different
and are also different from the one derived by the standard
computation with a large reheating temperature.

Let us start our analysis with the case of non-equilibrium
production and freeze-out. In this case, at early times the number
density $n_{\chi}$ is much smaller than $n_{\chi}^{eq}$ and the
Boltzmann equation (\ref{Ygeneral}) can be written as \be
\frac{dY}{dx} = 0.02095 \left(\frac{g_{\chi}}{g_{*s}}\right)^2
\frac{s}{H x} \langle \sigma^{ann}_{\chi}\ v \rangle~ x^{3}~
e^{-2x}.\label{YTlow}\ee Although this equation is valid only at
early times, it can be approximately integrated in the full range
of $x$, namely from $x=0$ to $x=\infty$ due to the exponential
suppression in the right hand side. Thus for the standard
cosmology (where $H \propto x^{-2}$), one can easily integrate
this equation and obtains \be Y_{\infty s} = 0.02095
\sqrt{\frac{\pi}{45}} g_{\chi}^2 g_*^{- 3/2} M_{pl}~ m_{\chi}
\left(\frac{a}{4} + 3 b \right).\label{noneq1}\ee The $Y_{\infty}$
is related to the mass density of $\chi$ particle today as
follows, at reheating we have $\rho_{\chi}(T_{RH}) = m_{\chi}
n_{\chi}(T_{RH}) = \frac{2 \pi^2}{45} g_{*s}(T_{RH}) m_{\chi}
Y_{\infty} T_{RH}^3$. After the reheating the universe is
radiation dominated and the following relation is satisfied
\cite{Giudice:2000ex}: \be
\frac{\rho_{\chi}(T_{now})}{\rho_{R}(T_{now})} =
\frac{T_{RH}}{T_{now}}~
\frac{\rho_{\chi}(T_{RH})}{\rho_R(T_{RH})}. \ee Therefore, in this
case $\Omega_{\chi} h^2$ is proportional to the annihilation cross
section instead of being inversely proportional as in case of high
reheating temperature.

Now we consider this scenario of low reheating with
non-equilibrium production and freeze-out in brane cosmology. The
Boltzmann equation is still given by Eq.(\ref{YTlow}), but with
$H= (k_2 x^{-4} + k_3 x^{-8})^{1/2}$ in the range of $x\in[0,x_t]$
and with the usual Hubble parameter $H= \sqrt{k_2} x^{-2}$ between
$x_t$ and $x=\infty$. Integrating this equation one finds \be
Y_{\infty b} \simeq 0.02095 \times 10^{-6} \sqrt{\frac{\pi}{45}}
g_{\chi}^2 g_*^{- 3/2} M_{pl}~ m_{\chi} \left( 9.46~  a + 37.8~ b
\right).\label{noneq2}\ee Here we have used $m_{\chi} \sim 100$
GeV and $M_5 \sim 10^6$ GeV, as an example, to do the integration
numerically. However, we have checked the result of $Y_{\infty b}$
for different values of $m_{\chi}$ and $M_5$. It turns out that
$Y_{\infty b}$ is diminished significantly for $M_5 \lsim 10^6$.
As can be observed from equations (\ref{noneq1}) and
(\ref{noneq2}) that this non-equilibrium scenario produces a very
suppressed relic density, particularly in brane cosmology.
Furthermore, the assumption that $n_{\chi} \ll n_{\chi}^{eq}$
impose a sever constraint on the annihilation cross section. Thus,
one can conclude that within this scenario, it is not possible to
account for the dark matter experimental results.

Now we turn to the second scenario in which the annihilation cross
section of the WIMP is large and hence it reaches the chemical
equilibrium before reheating. In this case, the computation of the
relic density $\Omega_{\chi} h^2$ is very close to the standard
case with high reheating temperature. At the early times \ie, when
$T > T_F$, the WIMP's are very close to equilibrium. So, as usual,
one can use the variable $\Delta(x) = Y(x) - Y^{eq}(x)$ to write
the Boltzmann equation as \be \Delta' = -(Y^{eq})^{'} - f(x)
\Delta \left(2 Y^{eq} + \Delta \right),\ee where $f(x)$ is given
by $ f(x) = \sqrt{\frac{\pi g_*}{45}} M_{pl} m_{\chi}
\left[\frac{a}{x^2} + \frac{6 b}{x^3}\right].$ Then by neglecting
$\Delta'$ and $\Delta^2$ , one obtains \be \Delta \simeq -
\frac{(Y^{eq})^{'}}{2 f(x) Y^{eq}}. \ee

At late time, when $T < T_F$ one gets $Y \gg Y^{eq}$ and hence we
can use the approximation $Y^2 - (Y^{eq})^2 \simeq Y^2$ in
Eq.(\ref{Boltzmann1}) and integrate from $T_F$ down to $T_{RH}$ to
determine $Y(T_{RH})$. In the standard cosmology,  one finds \be
\frac{1}{Y(x_{RH})}\Big\vert_{s} = \frac{1}{Y(x_F)} -
\sqrt{\frac{\pi g_*}{45}} M_{pl} m_{\chi} \left[\frac{a}{x} +
\frac{3 b}{x^2}\right]_{x_F}^{x_{RH}}, \label{eq4.7}\ee If it is
assumed that there is no entropy production for $T < T_{RH}$, then
there is no WIMP production for temperature below the reheating
temperature. Thus, the present value of $Y$ is given by
$Y(x_{RH})$ up to an overall correction due to the fact that the
reheating process is not complete at $T_{RH}$
\cite{Giudice:2000ex}. Here, two comments are in order: $i)$ As
mentioned above, $Y(x_F)^{-1} \simeq Y_{eq}(x_F)^{-1}$ which is of
order ${\cal O}(10^9)$, so its contribution to $Y(x_{RH})^{-1}$ in
Eq.(\ref{eq4.7}) can be neglected respected to the second term.
$ii)$ Since $T_{RH} < T_F$ (\ie, $x_{RH}
> x_F$), one can approximate Eq.(\ref{eq4.7}) and finds that the
relic abundance $\Omega_{\chi}h^2$ is very close to the one
obtained by using the standard calculation with high reheating
temperature, namely \be \Omega_{\chi} h^2 \sim 1.1 \times 10^{-11}
\left( \frac{a}{x_F} + \frac{3 b}{x_F^2} \right)^{-1},\ee which
implies that unless annihilation cross sections are quite small ($
\lsim 10^{-8}$), one gets, as usual, very small relic density.

In brane cosmology, Eq.(\ref{Boltzmann2}) describes the correct
Boltzmann equation that should be used. Also, as in the previous
scenario, one has to integrate this equation from $x_F$ to $x_t$
using brane cosmology feature, and from $x_t$ to $x_{RH}$ using
the standard cosmology feature. In this respect, one finds \bea
\frac{1}{Y(x_{RH})}\Big\vert_{b} &=&\frac{1}{Y(x_F)}-
\sqrt{\frac{\pi g_\ast}{45}}M_{pl} m_{\chi}
\left[3\sqrt{\frac{k_2}{k_3}} b
\left(\sinh^{-1}\left(\sqrt{\frac{k_3}{k_2}} x_F^{-2}\right) -
\sinh^{-1}\left(\sqrt{\frac{k_3}{k_2}}
x_t^{-2}\right)\right)\right. \nonumber\\
&+&\left. a \left(\frac{1}{x}~ _2F_1\left[\frac{1}{4},
\frac{1}{2}, \frac{5}{4}, \frac{-k_3}{k_2 x^4}
\right]\right)^{x_t} _{x_F} + \left(\frac{a}{x}+
\frac{3b}{x^2}\right)^{x_{RH}} _{x_t}\right].\label{Yb-2}\eea One
can easily check that the three parts of the second term in the
above equation are of the same order and give the dominant
contribution to $Y(x_{RH})^{-1}$. In this case, one can show that
the relic density (for $m_{\chi}=100$ GeV and $M_5=10^6$ GeV) is
given by \be \Omega_{\chi} h^2 \sim 1.1 \times 10^{-7} \left( 95.2
~ a - 4.12 ~ b\right)^{-1}.\ee From this equation it can be easily
seen that even with large annihilation cross section ${\cal
O}(10^{-6}-10^{-8})$, we are able to obtain the cosmologically
interesting value $\Omega_{\chi} h^2 \sim 0.1$. This implies that
the scenario of brane cosmology with low reheating termperature
and chemical equilibrium WIMPs is the most interesting model for
dark matter. It provide an interesting possibility for having dark
matter with large cross section (hence their detection would be
possible in future DM experiments) with suitable relic aboundanc.

\section{{\large \bf Conclusions}}
In this paper we have analyzed the relic abundance of cold dark
matter in brane cosmology. We have pointed out that the
consistency of the boundary conditions that account for the
presence of the brane implies that the $5D$ metric should contain,
at least, two warp factors. In case of one warp factor, we have
emphasized that the relation $\rho_b = - p_b$ is identically
satisfied, which contradicts the possibility of having matter on
the brane.

We have also studied the impact of brane cosmology on the cold
dark matter relic density. We investigated this effect in two
different scenarios, namely when the reheating temperature is
higher or lower than the freeze-out temperature. We showed that
with high reheating temperature, the relic density is enhanced
with many order of magnitude for $M_5 \lsim 10^{6}$. This imposes
one of the strongest constraints on the scale of large extra
dimensions. In case of low reheating temperature, we have
considered the possibility that WIMPs are in chemical equilibrium
or non-equilibrium, which depends on the value of their
annihilation cross section. We emphasized that if WIMPs are in
chemical non-equilibrium, then their relic density is very small
and they can not account for the observational limits. While in
case WIMPs reach  chemical equilibrium before reheating, we showed
that the relic density is enhanced by two order of magnitudes than
the standard thermal scenario result. This enhancement can be
considered as an interesting possibility for accommodating dark
matter with large cross section, which is favored by the detection
rate experiments.

\section*{{\large \bf Acknowledgements}}
We would like to thank O. Seto for useful discussion. A part of
this work was done within the Associate Scheme of ICTP.

\noindent


\end{document}